\documentclass[%
 reprint,
 amsmath,amssymb,
 aps,
 superscriptaddress,
 showkeys,
prb,
floatfix,
raggedbottom,
]{revtex4-2}

\usepackage{graphicx}
\usepackage{dcolumn}
\usepackage{bm}
\usepackage{hyperref}
\hypersetup{colorlinks, allcolors=blue}
\usepackage{physics}

\usepackage{siunitx}
\DeclareSIUnit\rydberg{Ry}

\begin{document}

\preprint{APS/123-QED}

\title{Disentangling Electron-Boson Interactions on the Surface of a Familiar Ferromagnet}%

\author{H\r{a}kon I. R\o{st}}
\affiliation{Department of Physics, Norwegian University of Science and Technology (NTNU), NO-7491 Trondheim, Norway.}
\affiliation{Department of Physics and Technology, University of Bergen, All\'egaten 55, 5007 Bergen, Norway.}

\author{Federico Mazzola}
\affiliation{Department of Molecular Sciences and Nanosystems, Ca’ Foscari University of Venice, 30172 Venice, Italy.}
\affiliation{Istituto Officina dei Materiali, Consiglio Nazionale delle Ricerche, Trieste I-34149, Italy.}

\author{Johannes Bakkelund}
\author{Anna Cecilie Åsland}
\affiliation{Department of Physics, Norwegian University of Science and Technology (NTNU), NO-7491 Trondheim, Norway.}

\author{Jinbang Hu}
\affiliation{Department of Physics, Norwegian University of Science and Technology (NTNU), NO-7491 Trondheim, Norway.}

\author{Simon P. Cooil}
\affiliation{Department of Physics and Centre for Materials Science and Nanotechnology, University of Oslo (UiO), 0318 Oslo, Norway.}

\author{Craig M. Polley}
\affiliation{MAX IV Laboratory, Lund University, Fotongatan 2, Lund, 22484 Sweden.}

\author{Justin W. Wells}
\email[Corresponding author: ]{j.w.wells@fys.uio.no}
\affiliation{Department of Physics, Norwegian University of Science and Technology (NTNU), NO-7491 Trondheim, Norway.}
\affiliation{Department of Physics and Centre for Materials Science and Nanotechnology, University of Oslo (UiO), 0318 Oslo, Norway.}

\begin{abstract}
 We report energy renormalizations from electron-phonon and electron-magnon interactions in spin minority surface resonances on Ni(111). The different interactions are identified, disentangled, and quantified from the characteristic signatures they provide to the complex self-energy and the largely different binding energies at which they occur. The observed electron-magnon interactions exhibit a strong dependence on momentum and energy band position in the bulk Brillouin zone. In contrast, electron-phonon interactions from the same bands appear to be relatively momentum- and symmetry-independent. Additionally, a moderately strong ($\lambda>0.5$) electron-phonon interaction is distinguished from a near-parabolic spin majority band not crossing the Fermi level.
\end{abstract}

\maketitle

\section{Introduction}\label{sec:intro}
In condensed matter, the interplay of electrons and other fundamental and collective excitations can induce new and exotic phases of electronic ordering. Perhaps most studied is the coupling between electrons and lattice vibrations (phonons) which can trigger an effective and attractive electron interaction and lead to superconductivity in elementary metals \cite{BCS1957theory}. While low $T_{\text{C}}$ superconductivity can be well explained from electron-phonon coupling (EPC) alone, other and less conventional pairing mechanisms have been suggested as ingredients of high $T_{\text{C}}$ superconductivity \cite{gweon2004unusual,inosov2011crossover,gotlieb2018revealing,oh2021evidence}. In superconducting ferro- and antiferromagnets, electrons can also couple to spin waves (magnons) \cite{saxena2000superconductivity,pfleiderer2001coexistence,aoki2001coexistence,curro2005unconventional,Kamihara2008iron,Karchev2003magnon,Zhang2011superconductivity}. Electron-magnon interactions are furthermore expected to mediate proximity-induced superconductivity across magnetic interfaces \cite{Kargarian2016amperean,Gong2017time,rohling2018superconductivity,thingstad2021eliashberg,Maeland2021electron,Ok2022strong}.

For the experimental study of many-body interactions with electrons, angle-resolved photoemission spectroscopy (ARPES) is the tool of choice as the complete, complex self-energy can be extracted from the measured electronic bandstructures \cite{kevan1992angle,damascelli2004probing,Hofmann2009electron,hufner2013photoelectron,hufner2007very}. While EPC has been extensively studied using ARPES, there are only a handful of reports of electron-magnon couplings (EMC) available \cite{Claessen2004,Cui2007high,Cui2007angle,Claessen2009,mlynczak2019kink,mazzola2022tuneable,andres2022strong,yu2022strong}. The majority of these consider couplings only in specific electron bands or over small sub-regions of reciprocal space. This motivates the need for further investigations of the EMC -- for instance, exploring how the interactions can vary between different spin bands and throughout the Brillouin zone (BZ).

Herein, we present a thorough study of the many-body interactions present in Ni(111). Electron-phonon and electron-magnon interactions are unraveled from different spin bands, at several different positions in the bulk BZ, and along several high-symmetry directions of the projected bulk BZ (hereinafter referred to as the PBZ). The EPC and EMC are disentangled from one another based on their characteristic signatures and contributions to the electron self-energy. The EMC of the spin minority bands exhibits a strong dependence on the electron momentum, i.e., the location within the bulk BZ, both in interaction strength and regarding the participating magnon modes. In contrast, the EPC is much less momentum-dependent and visible with reasonable strength in both the spin minority and majority bands.

First, an overview of the electronic structure of Ni(111) is given in Section~\ref{subsec:electronicStructure}. Next, the different many-body effects observed from its spin majority and minority bands are discussed, disentangled, and quantified in Section~\ref{sec:manyBody}. A summary and final remarks are given in Section~\ref{sec:conclusion}. Additional details about the experiments and many-body analysis are given in the Appendices \ref{sec:measurement}-\ref{sec:DFT} and the Supplementary Note \cite{Suppl_Mat}.

\section{Results and Discussion}
\subsection{Electronic Structure and Surface Resonances}\label{subsec:electronicStructure}
\begin{figure}[t!]
    \centering
    \includegraphics[]{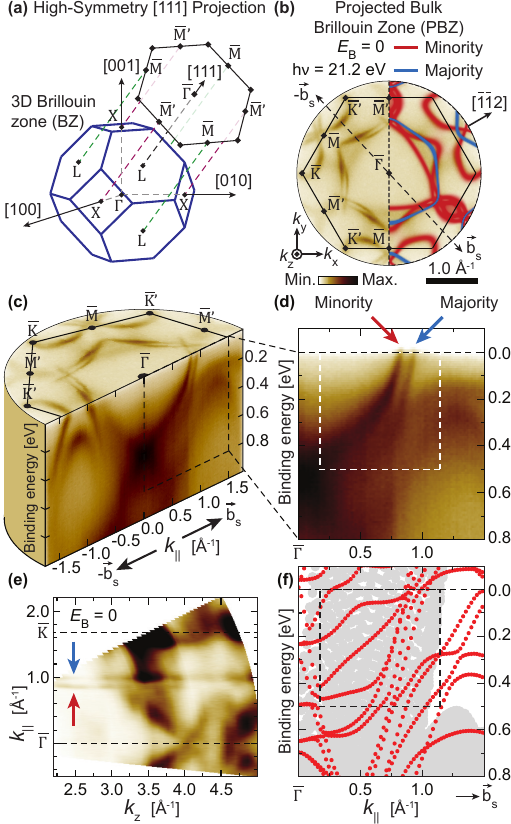}
    \vspace{-0.6cm}
    \caption{The electronic structure of Ni(111). \textbf{(a)}: Sketch showing the projection of the bulk Brillouin zone of Ni onto the (111) plane. \textbf{(b)}: Measured constant energy surface of Ni(111) at $E_{\text{F}}$ (left) and the calculated bandstructure with free-electron final states (right), both using $h\nu=\SI{21.2}{\eV}$.
    \textbf{(c)}: Volumetric representation of the measured Ni(111) bandstructure. The energy cut has been performed from $\bar{\Gamma}$ and along the $\pm\vb{b}_{\text{s}}$ directions as shown in (b). \textbf{(d)}: Measured bandstructure ($E$~vs.~$k_{||}$) along the $+\vb{b}_{\text{s}}$ direction. A clear spin splitting of the states can be seen close to $E_{\text{F}}$. The assignment of minority and majority states is based on the density functional theory calculation in (b). \textbf{(e)}: Photoemission intensity as a function of final state wave number $k_{\text{z}}$. \textbf{(f)}: The calculated, unrenormalized spin minority surface states (red) of Ni(111) along the $+\vb{b}_{\text{s}}$ direction. The shaded background (gray) represents the surface-projected, spin majority bulk bands. Surface resonance states can be observed near $(E_{\text{B}},k_{||})=(\SI{0.0}{\eV},0.82~\text{Å}^{-1})$.}
    \label{fig:fig1}
    \vspace{-0.6cm}
\end{figure}

An overview of the electronic structure of Ni(111) near the Fermi level ($E_{\text{F}}$) is shown in Fig.~\ref{fig:fig1}. Along the $\bqty{111}$ direction the bulk BZ of Ni is projected onto a two-dimensional zone that is hexagonal and threefold symmetric (Fig.~\ref{fig:fig1}a) \cite{Tserkezis2011photonic}.
The projected Fermi surface is shown in Fig.~\ref{fig:fig1}b: as measured by ARPES with $h\nu=\SI{21.2}{\eV}$ (left), and as calculated from first principles (DFT) while accounting for the available free-electron final-states at this photon energy (right) \cite{Liebowitz1978free,Aebi1994fermi,damascelli2004probing}.

Approximately halfway from $\Bar{\Gamma}$ towards the edge of the first PBZ, two different spin minority contours and one spin majority contour can be distinguished near $E_{\text{F}}$. The mentioned spin bands all meet near the  $\bar{\text{M}}'$ high-symmetry point, and approximately halfway between  $\bar{\text{M}}'$ and $\bar{\text{K}}$ a maximum separation between the bands is seen. Defining an in-plane momentum vector $\vb{b}_{\text{s}}$ ($\propto k_{||}$) from $\Bar{\Gamma}$ and towards the PBZ boundary between $\bar{\text{M}}'$ and $\bar{\text{K}}$ (Fig.~\ref{fig:fig1}b), the local energy dispersion $E(k_{||})$ of the spin bands can be investigated. An example cut along the directions $\pm\vb{b}_{\text{s}}$ is shown in Fig.~\ref{fig:fig1}c with two prominent spin bands highlighted (dashed rectangle). As further demonstrated in Fig.~\ref{fig:fig1}d, the two bands are nearly parallel and almost straight in the topmost \SI{100}{\milli\eV} near $E_{\text{F}}$. Based on the free-electron final-state dependent calculations (Fig.~\ref{fig:fig1}b), the dispersion furthest away from $\Bar{\Gamma}$ at $E_{\text{F}}$ is interpreted as a spin majority band, and the one closer to $\Bar{\Gamma}$ as one of two possible spin minority bands.

Notably, a two-dimensional, dispersionless behavior can be observed from the mentioned spin minority and majority bands. While their surrounding energy band features at $E_{\text{F}}$ readily disperse with wave vector $\vb{k}_{\text{z}}$ along the bulk $\Gamma\rightarrow\text{L}$ direction, the two parallel spin bands appear approximately linear, with little or no $\vb{k}_{\text{z}}$ dependence (Fig.~\ref{fig:fig1}e). These are common features of states that are localized or quasi-localized perpendicular at the atomic surface where the $\vb{k}_{\text{z}}$ symmetry is broken \cite{kevan1992angle}.

In Fig.~\ref{fig:fig1}f, the calculated, unrenormalized spin minority surface states of Ni(111) along $\vb{b}_{\text{s}}$ are shown, some of which appear to qualitatively resemble the dispersion of the measured spin minority states highlighted in Fig.~\ref{fig:fig1}d. The calculated spin minority states (Fig.~\ref{fig:fig1}f, red) appear to overlap with the projected spin majority states from the bulk (gray), thereby enabling coupling between surface and bulk states through spin-flip scattering processes by the absorption or emission of magnons \cite{Claessen2004}. A similar overlap is also present between the calculated spin majority surface states and the projected spin minority bulk bands (see the Supplementary Note~\cite{Suppl_Mat}). Hence the measured, $\vb{k}_{\text{z}}$-independent bands in Figs.~\ref{fig:fig1}d and \ref{fig:fig1}e are interpreted as surface resonance states \cite{Claessen2004,kevan1992angle}. The calculations suggest that a handful of similar states exist near $E_{\text{F}}$ with $k_{||}=0.7$-$0.8~\text{Å}^{-1}$. Several different resonances should, therefore -- in principle, be observable from ARPES measurements.

\subsection{Signatures of Electron-Boson Interactions}\label{sec:manyBody}
The spin bands in Figs.~\ref{fig:fig1}c and \ref{fig:fig1}d contain apparent signs of broadening over the binding energy range $E_{\text{B}}=0-\SI{400}{\meV}$. Typically, such energy broadening can be described by various quasiparticle renormalizations, signaling a reduced lifetime $\tau$ and associated increased self-energy $\Sigma$ for the electron states \cite{Hofmann2009electron}. One can often assume $\Sigma(\vb{k},\omega)$ to vary slowly with momentum when observed over a narrow range of $E$ vs. $\vb{k}$ -- for instance, within the interaction region of a renormalized electron energy band \cite{gayone2005determining}. The measured ARPES intensity at a temperature $T$ is then proportional to a simplified expression for the spectral function $\mathcal{A}(\vb{k},\omega)$ \cite{damascelli2004probing,Hofmann2009electron}:
\begin{align}\label{eq:spectral}
    &I(\vb{k},\omega)\propto\mathcal{A}(\vb{k},\omega) = \pi^{-1}\Im G(\vb{k},\omega) \nonumber \\ &=\frac{-\pi^{-1}\Im\Sigma(\omega)}{\bqty{\hbar\omega-\varepsilon\pqty{\vb{k}}-\Re\Sigma(\omega)}^2+\bqty{\Im\Sigma(\omega)}^2}.
\end{align}
$\mathcal{A}(\vb{k},\omega)$ is again proportional to the imaginary part of the one-particle Green's function for the photoexcitation process. Notably, Eq.~\ref{eq:spectral} states that for cuts through the data at constant binding energy $\hbar\omega$, $\mathcal{A}(\vb{k},\omega)$ will assume a Lorentzian line profile with a peak maximum at the value of $\vb{k}$ where $\hbar\omega=\varepsilon\pqty{\vb{k}}+\Re\Sigma(\omega)$ and a full width at half maximum $\text{FWHM}=2\Im\Sigma(\omega)$. By extracting and fitting momentum distribution curves (MDCs) at each measured energy one can then estimate the real and imaginary components of $\Sigma$ as \cite{gayone2005determining,Cui2007high,Claessen2009}:
\begin{align}
    \Re\Sigma\pqty{\omega} =& \,E\pqty{\vb{k}} - \varepsilon\pqty{\vb{k}}, \label{eq:ReSE} \\
    \vqty{\Im\Sigma\pqty{\omega}} =& \,\vqty{\dd \varepsilon/\dd \vb{k}}\cdot\vqty{\Delta \vb{k}}. \label{eq:ImSE}
\end{align}
Here, $E\pqty{\vb{k}}\equiv\hbar\omega_{\vb{k}}$ is the measured energy band dispersion, $\varepsilon\pqty{\vb{k}}$ is the unrenormalized band, $\vqty{\dd \varepsilon/\dd \vb{k}}$ is the absolute value of its gradient, and $\vqty{\Delta \vb{k}}$ is the measured peak half-width along the in-plane momentum ($k_{||}$) axis.

\begin{figure}
    \centering
    \includegraphics[]{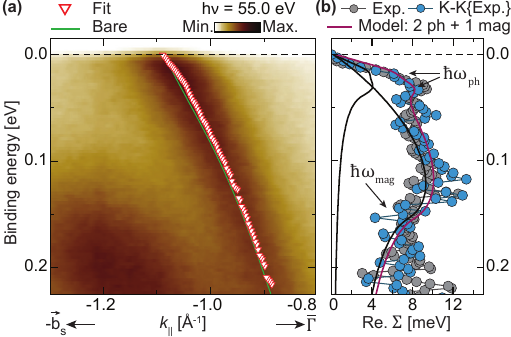}
    \caption{Electron-boson coupling (EBC) in the spin minority band along $-\vb{b}_{\text{s}}$ near the high-symmetry points $\bar{\text{M}}$ and $\bar{\text{K}}'$. \textbf{(a)}: The measured energy bandstructure, overlaid with the unrenormalized (green) band and the experimentally determined, renormalized spin band position (red triangles). \textbf{(b)}: The real self-energy $\Re\Sigma$ of the fitted band in (a). The $\Re\Sigma$ (gray) found from Eq.~\ref{eq:ReSE} is shown to satisfy causality with $\Im\Sigma$ through the Kramers-Kronig transformation (blue). A three-boson model (purple line) consisting of two distinct electron-phonon couplings (EPCs) with energies $\hbar\omega_{\text{ph}}=\SI{18}{\meV}$ and $\hbar\omega_{\text{ph}}=\SI{36}{\meV}$, respectively, and one electron-magnon coupling (EMC) with energy $\hbar\omega_{\text{mag}}=\SI{154}{\meV}$, best describes the measured line shape. The individual contributions from EPC and EMC (black lines) are also shown.}
    \label{fig:fig2}
\end{figure}

In Fig.~\ref{fig:fig2}a the fitted position of an interacting spin minority band of Ni(111) is shown. The same band that was also displayed along $-\vb{b}_{\text{s}}$ in Fig.~\ref{fig:fig1}c, but measured at a different $\vb{k}_{\text{z}}$ position in the bulk BZ. Several clear deviations between the fitted band position and the one-particle band structure can be seen within \SI{225}{\meV} of $E_{\text{F}}$, and sudden energy broadenings along the measured band position are also apparent. Both features are typical hallmarks of electron-boson coupling (EBC) \cite{hufner2007very,Claessen2004,Cui2007high,Cui2007angle,Claessen2009}. The experimental self-energy of the band was therefore estimated by a self-consistent analysis procedure based on Eqs.~\ref{eq:ReSE} and \ref{eq:ImSE} (see Appendix~\ref{sec:selfEnergy} for details). Its $\Re\Sigma$ is shown in Fig.~\ref{fig:fig2}b, and is demonstrated to satisfy causality with $\Im\Sigma$ via a Kramers-Kronig (K-K) transformation \cite{Kordyuk2005bare,Pletikosic2012finding}. From the spectrum a steep rise up to $\approx\SI{35}{\meV}$ can be observed, followed by a broad feature over a larger energy range. The former can be related to the apparent `kink' in the measured band position in Fig.~\ref{fig:fig2}a, and is a characteristic signature of EPC \cite{LaShell2000non,gayone2005determining,Kordyuk2005bare,Pletikosic2012finding,Higashiguchi2005energy}. Its energy is furthermore in approximate agreement with the Debye temperature of bulk Ni $\pqty{\Theta_{\text{D}}\approx\SI{477}{\kelvin}}$ \cite{stewart1983measurement}.

Deconvolving $\Re\Sigma$ by an integral inversion method produces the energy-dependent Eliashberg function $\alpha^{2}F\pqty{\omega}$, distinguishing the EBC modes that renormalize the locally measured electron band by their interaction energy \cite{jarrell1996bayesian,Gubernatis1991quantum,Shi2004direct,tang2004spectroscopic,chien2009anisotropic,chien2015electron}. The resultant $\alpha^{2}F\pqty{\omega}$ -- as detailed in the Supplementary Note \cite{Suppl_Mat}, immediately suggests two distinct couplings with energies matching the surface and bulk-derived vibrations of Ni \cite{Birgeneau1964normal,szeftel1985surface,stuhlmann1989surface,menezes1990surface}. In addition, it contains several prominent couplings above the Ni Debye energy $k_{\text{B}}\Theta_{\text{D}}$ and up to $E_{\text{B}}\approx\SI{140}{\meV}$ \cite{Suppl_Mat}.
 
Given the ferromagnetic nature of the system, interactions between electrons and magnons can occur \cite{Claessen2004,Cui2007high,Cui2007angle,Claessen2009}. At a glance, signatures of EMC from ARPES should crudely resemble those of EPC. However, the two will typically have separate coupling energies, and their functional shape should also differ because of their intrinsically different energy dispersion relations \cite{Suppl_Mat,mazzola2022tuneable}. Alternatively, phonon or magnon mode-specific renormalizations can -- under certain simplifying assumptions, be predicted from first principles calculations \cite{guistino2007lectron,Hellsing2018phonon,mahr2017implementation,He2023electron}. However, discovering all the possible couplings requires an overview of the occupied bosonic modes; a detailed knowledge of the EBC matrix elements; and all the possible propagation channels between initial and final states that preserve energy and momentum. This is a complicated task already in lighter elements \cite{chien2015electron,Hellsing2018phonon,He2023electron}, and an accurate determination is beyond the scope of this work. We therefore attempt only a preliminary assignment of coupling modes to our data, basing our suggestions on the distinct energies of the phonons and magnons in Ni.

Resultingly, the interactions at $E_{\text{B}}<k_{\text{B}}\Theta_{\text{D}}$ are assigned to EPC with surface- and bulk-derived phonons \cite{stuhlmann1989surface,Birgeneau1964normal,menezes1990surface}. Next, inelastic scattering measurements of bulk Ni have verified three different and characteristic magnon energy dispersions, namely: one acoustic branch along the $\bqty{111}$ propagation direction with energies $\hbar\omega_{\text{ac}}^{[111]}\leq\SI{175}{\milli\eV}$; another acoustic branch along $\bqty{100}$ with $\hbar\omega_{\text{ac}}^{[100]}\leq\SI{165}{\milli\eV}$; and one optical $\bqty{100}$ branch approximately between $130 \leq\hbar\omega_{\text{op}}^{[100]} \leq\SI{250}{\milli\eV}$ \cite{Mook1979neutron,Mook1985neutron,Mook1988temperature,Brookes2020spin}. The higher-energy signatures of $\alpha^{2}F\pqty{\omega}$ are thus interpreted as EMC primarily with the acoustic magnons. However, some features may stem from the optical branch near its lower-energy extremum.

To disentangle the interaction strengths of the EPC and EMC, a minimalistic three-boson approximation consisting of two distinct phonon modes and one magnon mode was fitted to the data from Fig.~\ref{fig:fig2}b. This model reproduced the main features of $\Re\Sigma$, with energies $\hbar\omega_{\text{ph}}^{(1)}=18\pm\SI{5}{\milli\eV}$, $\hbar\omega_{\text{ph}}^{(2)}=36\pm\SI{5}{\milli\eV}$, and $\hbar\omega_{\text{mag}}=154\pm\SI{6}{\milli\eV}$. The associated dimensionless EPC strength $\lambda_{\text{ph}}= 0.20\pm0.05$ is in excellent agreement with previous estimates from calculations and surface-sensitive inelastic scattering measurements \cite{Papaconstantopoulos1977calculations,holst2021material}. Furthermore, the EMC strength $\lambda_{\text{mag}}= 0.17\pm0.01$ approximately matches the value previously reported from ARPES \cite{Claessen2009}. The total EBC strength $\lambda_{\text{tot}}=\lambda_{\text{ph}}+\lambda_{\text{mag}}$ from the model is consistent within the value found from integral inversion analysis within the uncertainty (see Appendix~\ref{sec:selfEnergy}).

\begin{figure}
    \centering
    \includegraphics[]{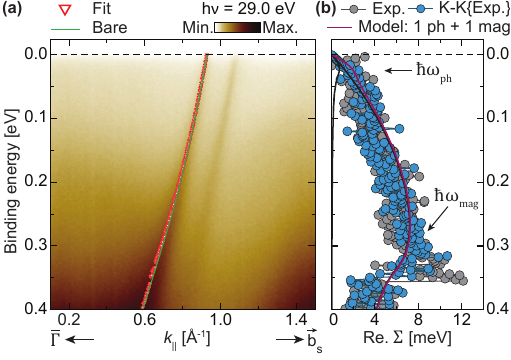}
    \caption{EBC in the spin-minority band along $+\vb{b}_{\text{s}}$ near the high-symmetry points $\bar{\text{M}}'$ and $\bar{\text{K}}$. \textbf{(a)}: The measured energy dispersion of the spin bands overlaid with the unrenormalized band (green) and the experimentally determined, renormalized spin minority band position (red triangles). \textbf{(b)}: The real self-energy $\Re\Sigma$ (gray) of the fitted band in (a), shown to be consistent with the imaginary part $\Im\Sigma$ through the K-K transformation (blue). A two-boson model (purple line) consisting of one EPC at $\hbar\omega_{\text{ph}}=\SI{23}{\meV}$; and one EMC at $\hbar\omega_{\text{mag}}=\SI{340}{\meV}$, best describes the measured line shape. The individual contributions from EPC and EMC (black lines) are also shown.}
    \label{fig:fig3}
\end{figure}

Interestingly, the innermost spin minority contour near the $\bar{\text{K}}$ and $\bar{\text{M}}'$ points revealed additional EBC modes with different characteristic energies and strengths \cite{Suppl_Mat}. This spin minority band, as measured and fitted along direction $+\vb{b}_{\text{s}}$, is shown in Fig.~\ref{fig:fig3} together with the real part of its K-K consistent self-energy. Coupling can be readily distinguished up to $E_{\text{B}}\approx\SI{350}{\meV}$, i.e., well beyond the maximum phonon and magnon energy values measured by inelastic scattering \cite{Birgeneau1964normal,Mook1985neutron}. However, calculations have indicated that the already mentioned optical magnon branch should exist in this energy range and out to the BZ boundary in $\vb{k}$ \cite{Cooke1985new}. EMC at similar energies in a different Ni spin minority band has also been reported \cite{Claessen2009}. We thus assign the higher-energy signatures to coupling primarily with optical magnons.

Similarly, the interactions as seen from Fig.~\ref{fig:fig3}b were quantified using a best-fit model consisting of two dominant boson modes: one EPC at $\hbar\omega_{\text{ph}}=23\pm\SI{12}{\milli\eV}$ with $\lambda_{\text{ph}}=0.05\pm0.03$; and one EMC at $\hbar\omega_{\text{mag}}=340\pm\SI{13}{\milli\eV}$ with $\lambda_{\text{mag}}=0.06\pm0.01$. The former is situated amidst the phonon energy range and suggests a weak coupling to either of the known vibrational modes \cite{Birgeneau1964normal,stuhlmann1989surface,menezes1990surface}. The latter indicates weak coupling to the optical magnons as already discussed \cite{Cooke1985new}. Extracting $\alpha^{2}F\pqty{\omega}$ by the integral inversion method yields interactions at similar energies and with a matching total EBC strength $\lambda_{\text{tot}}$ \cite{Suppl_Mat}. The reason behind the different participating boson modes and the smaller $\lambda_{\text{tot}}$ when compared to the EBC measured along $-\vb{b}_{\text{s}}$ (Fig.~\ref{fig:fig2}) is not immediately clear. Possibly, it may originate from having a different pairing of suitable initial and final electron states available at this position in the bulk BZ \cite{hofmann2006surfaces}.

The estimated coupling energies and strengths $\lambda$ from the two different $E$~\text{vs.}~$k_{||}$ cuts along $\pm\vb{b}_{\text{s}}$ have been summarized in Table~\ref{tab:Couplings}. Coupling parameters for the same cuts but measured at additional $\vb{k}_{\text{z}}$ positions in the BZ, i.e. using different photoexcitation energies $h\nu$, have also been presented. Their corresponding plots are shown and discussed in the Supplementary Note \cite{Suppl_Mat}.

\begin{figure*}
    \centering
    \includegraphics[]{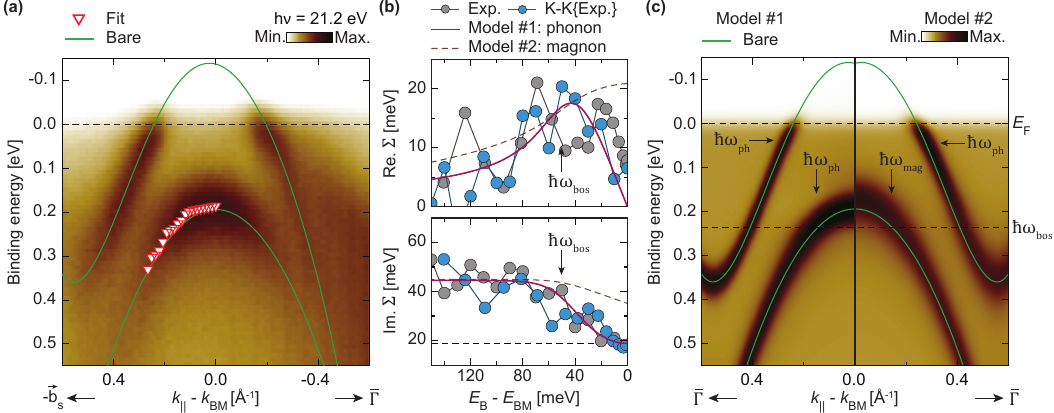}
    \caption{EBC in the spin majority energy band along $-\vb{b}_{\text{s}}$ near the high-symmetry points $\bar{\text{M}}$ and $\bar{\text{K}}'$. \textbf{(a)}: The measured electron energy dispersion along $-\vb{b}_{\text{s}}$, overlaid with the experimentally determined, renormalized position of the spin majority band (red triangles), and suggested one-particle `bare' bands for both spin configurations (in green). \textbf{(b)}: The real and imaginary self-energies of the fitted spin majority band in (a). Each component is shown to be consistent through the K-K transformation. The EBC appears at $\SI{50}{\milli\eV}$ below the spin majority energy band maximum $E_{\text{BM}}$. The interaction is best described by EPC from $E_{\text{BM}}$ with $\lambda_{\text{ph}}=0.55$ (purple) instead of EMC from $E_{\text{F}}$ (dashed gray). The added energy broadening from electron-impurity scattering and the finite instrumental resolution (dashed horizontal black line) is also shown. \textbf{(c)}: ARPES simulations of the spin bands in (a), implementing either EMC at $\hbar\omega_{\text{mag}}=\SI{235}{\meV}$ below $E_{\text{F}}$ (right) or EPC at $\hbar\omega_{\text{ph}}=\SI{50}{\meV}$ below $E_{\text{BM}}$ (left) in the spin majority band. Both models have an additional EPC contribution in the spin minority band at $\hbar\omega_{\text{ph}}=\SI{35}{\meV}$ below $E_{\text{F}}$, as suggested from the self-energy analysis summarized in Table~\ref{tab:Couplings}.}
    \label{fig:fig4}
\end{figure*}

Finally, attention is directed towards electron-boson interactions in the spin majority states. Close to midway between $\bar{\text{M}}$ and $\bar{\text{K}}'$ a near parabolic spin majority band is found at $\SI{185}{\milli\eV}$ below $E_{\text{F}}$. This band also appeared along $-\vb{b}_{\text{s}}$ in Fig.~\ref{fig:fig2}a and was found to reach a global band maximum when measured with $h\nu=\SI{21.2}{\eV}$ \cite{Suppl_Mat}. In Fig.~\ref{fig:fig4}a, the fitted spin majority band position is overlaid on the measured bandstructure, together with the one-particle band suggested by K-K analysis. To obtain a more accurate estimate of the band maximum, the renormalized band was asserted from a combination of MDC and energy distribution curve (EDC) fits \cite{LaShell2000non,Hofmann2009electron}.

\begin{table*}[t]
    \centering
    \vspace{-0.6cm}
    \caption{Measured EBC in the spin minority and majority band(s) as a function of photoexcitation energy $h\nu$. Entries marked with a `$*$' are shown and discussed in the Supplementary Note \cite{Suppl_Mat}. The mass-enhancement factors $\lambda_{\text{tot}}$ and the constituents $\lambda_{\text{ph}}$ and $\lambda_{\text{mag}}$ were estimated separately using different methods (see Section~\ref{sec:selfEnergy}). The uncertainties in $\lambda_{\text{tot}}$ were obtained by propagating the uncertainty of $\alpha^{2}F\pqty{\omega}$ as extracted from the data \cite{jarrell1996bayesian}. All other uncertainties were estimated from a relative $5\%$ increase in the root mean square (RMS) difference between the corresponding best-fit EBC model and the data.}
    \vspace{0.3cm}
    \label{tab:Couplings}
    \begin{tabular}{llccccccc}
        \hline
        \hline
        Figure$\quad\,\,$ & Fitted band dispersion$\quad$ & $h\nu$~[eV]$\quad$ & $\quad\lambda_{\text{tot}}\quad$ & $\quad\hbar\omega_{\text{ph}}$~[meV]$\quad$ & $\lambda_{\text{ph}}\quad$ & $\quad\hbar\omega_{\text{mag}}$~[meV]$\quad$ & $\lambda_{\text{mag}}\quad$ & Temp.~[K] \\
        \hline
        Fig.~\ref{fig:fig2} & Min. spin $\Bar{\Gamma}\rightarrow-\vb{b}_{\text{s}}\quad$ & $55.0\quad$ & $0.24\pm0.06$ & $18,36\pm 5$ & $0.20 \pm 0.05$ & $154 \pm 6\,\,\,$ & $0.17 \pm 0.01$ & $\,\,\,21$ \\
        Fig.~S6$^{*}$ & Min. spin $\Bar{\Gamma}\rightarrow-\vb{b}_{\text{s}}\quad$ & $29.0\quad$ & $0.14\pm0.06$ & $\quad\,\,\,32\pm2$ & $0.13 \pm 0.02$ & $\,\,\,-$ & $-$ & $\,\,\,77$ \\
        Fig.~\ref{fig:fig3} & Min. spin $\Bar{\Gamma}\rightarrow+\vb{b}_{\text{s}}\quad$ & $29.0\quad$ & $0.08\pm 0.05$ & $\quad\quad23\pm12$ & $0.05 \pm 0.03$ & $340 \pm 13$ & $0.06 \pm 0.01$ & $\,\,\,77$ \\
        Fig.~S7$^{*}$ & Min. spin $\Bar{\Gamma}\rightarrow+\vb{b}_{\text{s}}\quad$ & $21.2\quad$ & $0.20\pm0.07$ & $18,30 \pm 8$ & $0.12 \pm 0.04$ & $250 \pm 6\,\,\,$ & $0.11 \pm 0.01$ & $115$ \\
        Fig.~\ref{fig:fig4} & Maj. spin $\Bar{\Gamma}\rightarrow-\vb{b}_{\text{s}}\quad$ & $21.2\quad$ & $0.60\pm0.32$ & $\quad\,\,\,50\pm5$ & $0.55 \pm 0.05$ & $\,\,\,-$ & $-$ & $115$ \\
         \hline
         \hline
    \end{tabular}
\end{table*}

The resultant self-energy $\Sigma$ (Fig.~\ref{fig:fig4}b) indicates an EBC at $E_{\text{B}}=\SI{235}{\milli\eV}$, i.e., approximately $\SI{50}{\milli\eV}$ below the band maximum energy $E_{\text{BM}}$. This is similar to one of the EMC energies reported from the spin minority bands (see Table~\ref{tab:Couplings}). However, EMC at this energy does not satisfy the line shape of the measured $\Sigma$. Specifically, it fails to reproduce the abrupt `step' that is observed from $\Im\Sigma$. In comparison, a much better fit can be achieved using an EPC model that allows coupling from $E_{\text{BM}}$, with $\hbar\omega_{\text{ph}}=50\pm\SI{5}{\milli\eV}$ and $\lambda_{\text{ph}}=0.55\pm0.05$. The abrupt increase in the electron density of states (DOS) at the band maximum then causes a relatively large $\lambda_{\text{ph}}$ compared to the EPC found near $E_{\text{F}}$. Similarly strong EPC in near-parabolic bands below $E_{\text{F}}$ has been reported previously from ARPES, but only in two-dimensional and non-magnetic materials \cite{Mazzola2013kinks,Mazzola2017strong,rost2023phonon}.

The data shown in Figs.~\ref{fig:fig4}a and \ref{fig:fig4}b alone cannot give one definite answer about the bosonic origin of the interaction observed from the spin majority band. However, EPC appears to be more likely, based on the line shape of the $\Sigma$ contributions. To further explore the origin of the observed EBC, the spectral function $\mathcal{A}(\vb{k},\omega)$ from Fig.~\ref{fig:fig4}a was simulated within either of the two suggested coupling schemes. The simulations were performed using the suggested one-particle bands shown (i.e. `bare' bands, in green), broadened by the experimental resolutions, and a Fermi-Dirac distribution at the measurement temperature $T$ (numbers in Appendix~\ref{sec:measurement}). Based on additional measurements of the adjacent spin minority band (see Table~\ref{tab:Couplings}), an EPC near $E_{\text{F}}$ with $\hbar\omega_{\text{ph}}=\SI{35}{\milli\eV}$ and $\lambda_{\text{ph}}=0.1$ was also included. The two simulations are shown side-by-side in Fig.~\ref{fig:fig4}c. Within the topmost \SI{100}{\milli\eV} below $E_{\text{F}}$ the two suggestions are very similar, as both have the same, dominant $\Sigma$ contribution from EPC in the spin minority band at $E_{\text{B}}=\hbar\omega_{\text{ph}}=\SI{35}{\milli\eV}$. At larger binding energies, however, they reveal more striking differences. The suggested EMC significantly renormalizes the near-parabolic spin majority states, shifting these away from the one-particle energy band maximum and towards $E_{\text{F}}$. A strong energy broadening is also present throughout the simulated energy range. In comparison, EPC within the spin majority band yields a more local renormalization: here, the energy broadening is concentrated around the one-particle energy band maximum at $E_{\text{BM}}$ and a more pronounced kinking of the bandstructure is observed.

The simulated EPC is thus seen to better recreate the measured shape of the spin majority band (Fig.~\ref{fig:fig4}a) and its associated self-energy $\Sigma$ (Fig.~\ref{fig:fig4}b). The similar, but weaker coupling observed from the spin minority bands (Table~\ref{tab:Couplings}) certainly confirms the presence of appreciable EPC in the system, which should also occur in the spin majority bands. Furthermore, one could argue that intra-band scattering from electron-phonon interactions can -- in many cases, occur with a higher probability than inter-band electron-magnon scattering.

\section{Conclusions}\label{sec:conclusion}
In summary, we have demonstrated the existence and $\vb{k}_{\text{z}}$-dependence of both phonon- and magnon-derived quasiparticles in spin-minority, surface resonance energy bands on Ni(111). These have been disentangled based on their characteristic interaction energies and the functional form of their self-energy contributions. Different electron-magnon interactions have been observed and assigned to the distinctly different magnon modes available in the system. Previously unanticipated acoustic mode coupling has been demonstrated, and higher-energy optical mode coupling has been re-affirmed and re-interpreted, adding rigor to previous works \cite{Mook1979neutron,Mook1985neutron,Mook1988temperature,Brookes2020spin,Cooke1985new,Claessen2009}. The specific magnon mode that electrons interact with -- and also their associated coupling strengths $\lambda_{\text{mag}}$, have been shown to vary dramatically with the spin minority band position within the bulk BZ.

Additionally, a moderately strong ($\lambda>0.5$) renormalization has been observed in bulk spin majority bands at larger binding energies. This feature is best described by electron-phonon coupling near the corresponding energy band maximum and is -- to the best of our knowledge, the first known reporting of such couplings from spin-polarized, sub-Fermi level energy band maxima in three-dimensional ferromagnets.

\section{Acknowledgements}
This work was partly supported by the Research Council of Norway (RCN), project numbers 324183, 315330, and 262633. Additional financial support was received from CALIPSOplus, under Grant Agreement 730872 from the EU Framework Programme for Research and Innovation HORIZON 2020. We acknowledge Elettra Sincrotrone Trieste and MAX IV Laboratory for providing time on their beamlines APE-LE and Bloch, respectively, and for technical support. Research conducted at MAX IV, a Swedish national user facility, is supported by the Swedish Research Council under contract 2018-07152, the Swedish Governmental Agency for Innovation Systems under contract 2018-04969, and Formas under contract 2019-02496. We would also like to thank T. Balasubramanian, T.-Y. Chien, B. Holst, R. Manson, J. Shi, A. Skarpeid, A. Sudbø, and E. Thingstad for insightful discussions on electron-boson interactions.

\section{Methods}\label{sec:methods}

\subsection{Sample Preparation \& Bandstructure Measurements}\label{sec:measurement}
A clean Ni(111) surface was prepared by subjecting a bulk crystal to repeated cycles of Ar$^+$ ion sputtering at \SI{1}{\kilo\eV}, followed by annealing to \SI{500}{\celsius} for a short duration. The cleanliness of the surface was verified using low-energy electron diffraction (LEED) and X-ray photoelectron spectroscopy (XPS) of the relevant core levels (see the Supplementary Note \cite{Suppl_Mat}).

Energy bandstructure measurements were performed at \SI{115}{\kelvin} using a NanoESCA III aberration-corrected EF-PEEM equipped with a He I photoexcitation source ($h\nu=\SI{21.2}{\eV}$), using pass energy $E_{\text{P}}=\SI{25}{\eV}$ and a \SI{0.5}{\mm} entrance slit to the energy filter \cite{Escher2005NanoESCA}. At the mentioned settings, the instrument had $E$ and $k$ resolutions of approximately \SI{50}{\meV} and $0.02~\text{Å}^{-1}$, respectively. 

Higher energy resolution bandstructure measurements were performed at the synchrotron endstations APE-LE (Elettra, Trieste, Italy) and Bloch (MAX IV Laboratory, Lund, Sweden). At Elettra the Ni(111) crystal was cooled to $T=\SI{77}{\kelvin}$ and measured with an energy resolution $\Delta E=\SI{12}{\milli\eV}$.
At Bloch the crystal temperature was $T=\SI{21}{\kelvin}$ and the energy resolution $\Delta E\leq\SI{8}{\milli\eV}$. 
All measurements at both facilities were performed using VG SCIENTA DA30 analyzers.

\subsection{Self-Energy $\Sigma$ Analysis}\label{sec:selfEnergy}
Momentum distribution curves (MDCs) of the spin majority and minority bands were fitted over the relevant energy ranges using one or more Lorentzian line shapes, superimposed on a linear or polynomial background. The peak position and width (in $\text{Å}^{-1}$) of the bands at each measured energy were extracted, and in turn used to estimate $\Re\Sigma$ and $\Im\Sigma$, respectively. An initial guess at the one-particle band was approximated by a $5^\text{th}$ degree polynomial over the same energy range as the fitted experimental data. Its shape was then adjusted to achieve causality between the self-energy $\Sigma$ components via the Kramers-Kronig (K-K) transform \cite{Kordyuk2005bare,Pletikosic2012finding,Mazzola2017strong,mazzola2022tuneable}.

To disentangle any bosonic contributions to the measured self-energies, the Eliashberg coupling function $\alpha^{2}F\pqty{\omega}$ was extracted from each $\Sigma\pqty{\omega,T}$ by an integral inversion method \cite{jarrell1996bayesian,Gubernatis1991quantum,Shi2004direct,tang2004spectroscopic,chien2009anisotropic,chien2015electron}. The constant energy offset $\delta E$ of each $\Im\Sigma$ was assigned to broadening from a combination of electron-impurity scattering and the finite instrumental resolution \cite{Valla1999many}. Data points from the corresponding $\Re\Sigma$ at energies $E_{\text{B}}<\delta E$ were discarded to correct for distortions of the renormalized band position near $E_{\text{F}}$ \cite{chien2015electron}. Any signs of electron-electron scattering were indiscernible within the energy ranges measured and therefore excluded \cite{echenique2004decay,chulkov2003hole}. From each extracted $\alpha^{2}F\pqty{\omega}$, the total electron mass-enhancement $\lambda_{\text{tot}}$ due to EBC in the quasielastic approximation was estimated as \cite{grimvall1981electron}:
\begin{equation}\label{eq:lambda}
    \lambda_{\text{tot}} = 2\int_{0}^{\omega_{\text{max}}}\frac{\alpha^{2}F\pqty{\omega'}}{\omega'}\dd\omega',
\end{equation}
where $\omega_{\text{max}}$ is the frequency of the observable bosonic mode with the highest energy.

To further quantify the individual EBC contributions to the measured $\Sigma$, linear contributions of EPCs and EMCs were simulated and compared to the line shapes found from the experimental data. Each measured $\Sigma$ was fitted individually to minimize the root mean square (RMS) difference to the simulation of postulated couplings at the experimental temperature $T$. The phonon occupancy was approximated by a three-dimensional Debye model with phonon DOS $\rho_{\text{ph}}\pqty{\omega}\propto\omega^{2}$ and Debye frequency $\omega_{\text{max}}^{\text{ph}}$ \cite{Hellsing2002electron}. A similar model was used for the magnons, with a maximum frequency $\omega_{\text{max}}^{\text{mag}}$ and a magnon DOS $\rho_{\text{mag}}\pqty{\omega}\propto\omega^{1/2}$, based on the energy dispersion $\omega\propto\vb{q}^{2}$ expected for acoustic magnons \cite{Claessen2004,Claessen2009}.

Each linear contribution ($i$) to $\Im\Sigma$ from EBC was calculated as \cite{gayone2005determining,Valla1999many}:
\begin{align}\label{eq:Eliashberg}
    \Im\Sigma^{(i)}\pqty{\omega,T} =\,\, &\pi\hbar\int_{0}^{\omega_{\text{max}}}\alpha^{2}F^{(i)}\pqty{\omega'}\cdot[1+2n\pqty{\omega',T} \nonumber \\
    &+f\pqty{\omega+\omega',T}-f\pqty{\omega-\omega',T}]\dd\omega',
\end{align}
where $\alpha^{2}F^{(i)}\pqty{\omega}$ is the Eliashberg coupling function for the interaction $i$, and $n\pqty{\omega,T}$ and $f\pqty{\omega,T}$ are boson and fermion distributions, respectively. For phonons in the isotropic Debye model, $\alpha^{2}F^{(i)}\pqty{\omega}=\lambda_{\text{ph}}\pqty{\omega/\omega_{\text{max}}^{\text{ph}}}^{2}$ when $\omega<\omega_{\text{max}}^{\text{ph}}$ is assumed, with $\lambda_{\text{ph}}$ being the dimensionless strength of the EPC \cite{LaShell2000non}. For isotropic magnons, $\alpha^{2}F^{(i)}\pqty{\omega}=\pqty{\lambda_{\text{mag}}/4}\pqty{\omega/\omega_{\text{max}}^{\text{mag}}}^{1/2}$ for energies $\omega<\omega_{\text{max}}^{\text{mag}}$, with $\lambda_{\text{mag}}$ being the dimensionless EMC strength \cite{Claessen2004}. In each case, $\alpha^{2}F^{(i)}\pqty{\omega}=0$ above $\omega_{\text{max}}$. The $\Re\Sigma^{(i)}$ corresponding to each $\Im\Sigma^{(i)}$ was found using the K-K transform \cite{Kordyuk2005bare,Pletikosic2012finding}.

\subsection{First Principles Calculations}\label{sec:DFT}
First-principles calculations were performed using the density functional theory (DFT) software package QuantumESPRESSO. A plane wave basis with ultra-soft pseudopotentials and the local density approximation (LDA) for the exchange-correlation energy were used. Bulk calculations were performed self-consistently in a system with periodic boundary conditions, and $k$ points were sampled using a Monkhorst-Pack grid of $12\times12\times12$. The cut-off energy was \SI{50}{\rydberg} and the convergence threshold \SI{1e-8}{\rydberg}. The surface states were calculated using a slab geometry with $24$ atomic layers and a separation of \SI{15}{\angstrom} between slabs. Sampling of $k$ points was done using a Monkhorst-Pack grid of $10\times10\times1$. The cut-off energy was \SI{40}{\rydberg} and the convergence threshold \SI{1e-6}{\rydberg}.

To simulate the ARPES spectra visible when using photoexcitation energy of $h\nu=\SI{21.2}{\eV}$, the crystal momentum of the emitted electrons was selected according to the free-electron final state approximation of photoemission \cite{damascelli2004probing,Osterwalder2006}. In the extended BZ, a hemispherical cut with a radius $|\vb{k}_{\text{F}}|$ corresponding to the free-electron final state wave vector at $E_{\text{F}}$ was calculated around $\Gamma$. Only the states at the Fermi surface coinciding with the hemispherical shell were projected onto the (111) plane. See for instance Refs.~\onlinecite{Aebi1994fermi} and \onlinecite{Osterwalder1996} for additional details on the methodology. In determining the sphere radius, a work function $\phi_{\text{s}}=\SI{4.9}{\eV}$ (found by measurement, see the Supplementary Note \cite{Suppl_Mat}) and the value $V_{0}=\SI{10.7}{\eV}$ \cite{Osterwalder1996} for the inner potential was assumed. The corresponding projected bulk states yielded the spin-polarized constant energy surface displayed in Fig.~\ref{fig:fig1}b, matching the positions of the measured bands within an 8\% difference. We assign this error to uncertainties in the $\phi_{\text{s}}$ and $V_{0}$ values used to determine the radius $|\vb{k}_{\text{F}}|$ of the hemisphere.

%

\end{document}


\title{Supplementary Note:\\Disentangling Electron-Boson Interactions on the Surface of a Familiar Ferromagnet}%

\author{H\r{a}kon I. R\o{st}}
\affiliation{\footnotesize{Department of Physics, Norwegian University of Science and Technology (NTNU), NO-7491 Trondheim, Norway.}}
\affiliation{\footnotesize{Department of Physics and Technology, University of Bergen, All\'egaten 55, 5007 Bergen, Norway.}}

\author{Federico Mazzola}
\affiliation{\footnotesize{Department of Molecular Sciences and Nanosystems, Ca’ Foscari University of Venice, 30172 Venice, Italy.}}
\affiliation{\footnotesize{Istituto Officina dei Materiali, Consiglio Nazionale delle Ricerche, Trieste I-34149, Italy.}}

\author{Johannes Bakkelund}
\author{Anna Cecilie Åsland}
\affiliation{\footnotesize{Department of Physics, Norwegian University of Science and Technology (NTNU), NO-7491 Trondheim, Norway.}}

\author{Jinbang Hu}
\affiliation{\footnotesize{Department of Physics, Norwegian University of Science and Technology (NTNU), NO-7491 Trondheim, Norway.}}

\author{Simon P. Cooil}
\affiliation{\footnotesize{Department of Physics and Centre for Materials Science and Nanotechnology, University of Oslo (UiO), 0318 Oslo, Norway.}}

\author{Craig M. Polley}
\affiliation{\footnotesize{MAX IV Laboratory, Lund University, Fotongatan 2, Lund, 22484 Sweden.}}

\author{Justin W. Wells}
\email[Corresponding author: ]{j.w.wells@fys.uio.no}
\affiliation{\footnotesize{Department of Physics, Norwegian University of Science and Technology (NTNU), NO-7491 Trondheim, Norway.}}
\affiliation{\footnotesize{Department of Physics and Centre for Materials Science and Nanotechnology, University of Oslo (UiO), 0318 Oslo, Norway.}}

\maketitle
\vspace{-0.2cm}
\section*{Surface cleanliness and work function}
\begin{figure}[t]
    \centering
    \includegraphics[]{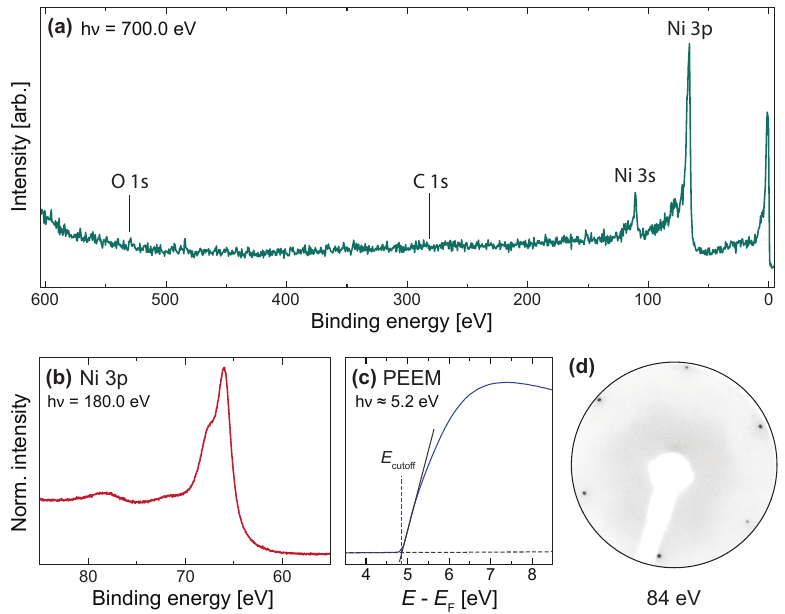}
    \caption{Photoelectron spectroscopy and surface diffraction measurements of clean Ni(111). \textbf{(a)}: Survey XPS scan of Ni(111). The Ni core levels $3s$ and $3p$ are visible, while any signatures from typical contaminants or adsorbates (C~$1s$, O~$1s$) are negligible. \textbf{(b)}: Higher resolution, narrow XPS scan of the Ni~$3p$ core level. \textbf{(c)}: The low-energy secondary electron cutoff (SEC) from Ni(111) measured using PEEM. The extracted value for the cutoff $E_{\text{cutoff}}=\SI{4.9}{\eV}$ equals the surface work function $\phi_{\text{S}}$. \textbf{(d)}: Low-energy electron diffraction (LEED) pattern after the preparation of a clean Ni(111) surface.}
    \label{fig:figS1}
\end{figure}

Prior to the energy bandstructure measurements, the cleanliness of the Ni(111) was asserted using surface-sensitive core level spectroscopy and diffraction as shown in Fig.~\ref{fig:figS1}. The sample was repeatedly sputtered using Ar$^+$ ions and subsequently annealed in cycles until no C~$1s$ nor O~$1s$ signatures could be detected. The crystalline quality of the surface was verified using low-energy electron diffraction (LEED).

The surface work function $\phi_{\text{S}}$ was obtained by measuring the low-energy secondary electron cutoff (SEC) from photoemission shown in Fig.~~\ref{fig:figS1}c. The SEC measurements were performed using a NanoESCA III aberration-corrected EF-PEEM equipped with a Hg photoexcitation source ($h\nu\approx\SI{5.2}{\eV}$). The measurements were performed with a pass energy $E_{\text{P}}=\SI{50}{\eV}$ and a \SI{0.5}{\milli\metre} entrance slit, resulting in an energy resolution of $\approx\SI{100}{\milli\eV}$. In the NanoESCA setup the energy of emitted photoelectrons was measured relative to the Fermi level, i.e., $E-E_{\text{F}}$:
\begin{equation}
    E-E_{\text{F}} = E_{\text{K}} + \phi_{\text{S}},
\end{equation}
with $E_{\text{K}}$ being their kinetic energies leaving the Ni(111) surface. For $E_{\text{K}}<\phi_{\text{S}}$ the electrons were unable to escape, and the measured energy of the SEC was therefore found to be $E_{\text{cutoff}}=\phi_{\text{S}}$ \cite{Escher2005NanoESCA}.

\section*{Bulk spin minority and majority states at the Fermi level}
\begin{figure}
    \centering
    \includegraphics[]{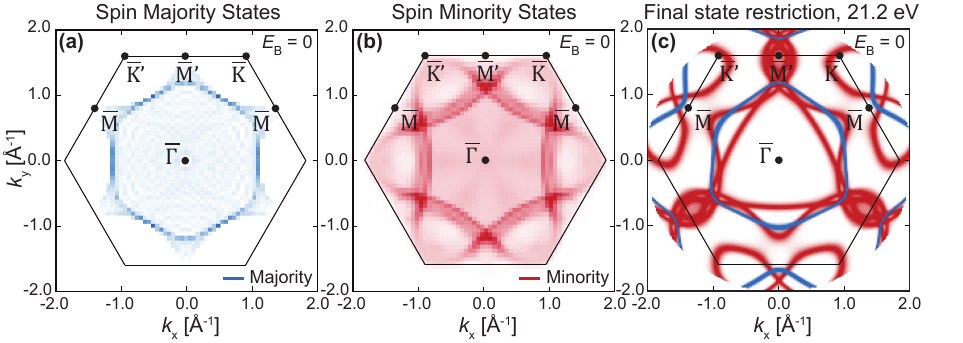}
    \caption{The spin bands of Ni(111) at the Fermi level ($E_{\text{B}}=0$) as calculated from first principles.\\ \textbf{(a,b)}: The majority (a) and minority (b) spin states within the $1^{\text{st}}$ bulk BZ of Ni, projected onto the (111) plane. The outline of the  $1^{\text{st}}$ projected bulk BZ (PBZ) has been overlaid. \textbf{(c)}: The (111)-projected spin majority and minority states of Ni that satisfy the free-electron final-state approximation ($h\nu=\SI{21.2}{\eV}$; $\phi_{\text{s}}=\SI{4.9}{\eV}$; $V_{0}=\SI{10.7}{\eV}$) \cite{Osterwalder1996}.}
    \label{fig:figS2}
\end{figure}

Projections of the Fermi level spin majority and minority states from bulk Ni and onto the (111) plane are shown in Figs.~\ref{fig:figS2}a and \ref{fig:figS2}b, respectively. The states as depicted have been calculated using density functional theory (DFT), summing over all out-of-plane (i.e. $\vb{k}_{\text{z}}$ momenta) within the first Brillouin zone (BZ). They have also been weighted by a Fermi-Dirac distribution to account for the thermal broadening at $T=\SI{115}{\kelvin}$, i.e., the sample temperature during the corresponding constant energy surface bandstructure measurements (see Appendix V A in the main text).

The initial electronic states that are visible from photoemission can be evoked by considering only the available free-electron final states that satisfy the conservation of energy and crystal momentum \cite{Aebi1994fermi,Osterwalder1996,damascelli2004probing}. A hemisphere is drawn around $\Gamma$ in the extended BZ scheme with a radius $|\vb{k}_{\text{F}}|$, i.e, the magnitude of the free-electron final state wave vector defined by the photoexcitation energy. Only the bound states coinciding with this hemisphere will then contribute to the observable photoemission intensity. An example is shown in Fig.~\ref{fig:figS2}c, where the Fermi level states that are visible using $h\nu=\SI{21.2}{\eV}$ have been projected onto the (111) plane. The resultant spin minority and majority states faithfully reproduce the measured Fermi level energy surface of Ni(111), as shown and discussed in Section~II~A of the main text.

\section*{Surface spin minority and majority resonance states}
Along $\Gamma\rightarrow\text{L}$ in the bulk BZ of Ni, i.e., the direction normal to the (111) plane, a subset of the measured spin minority and majority states show no apparent energy dispersion with out-of-plane electron momentum $\hbar\vb{k}_{\text{z}}$ (Fig.~1e, main text). This behavior is typical for states that are quasi-localized at the atomic surface, for instance, resonances between bulk ($\vb{k}_{\text{z}}$-dependent) and surface ($\vb{k}_{\text{z}}$-independent) spin states \cite{kevan1992angle,Claessen2004}.

\begin{figure}[ht]
    \centering
    \vspace*{-5mm}
    \includegraphics[]{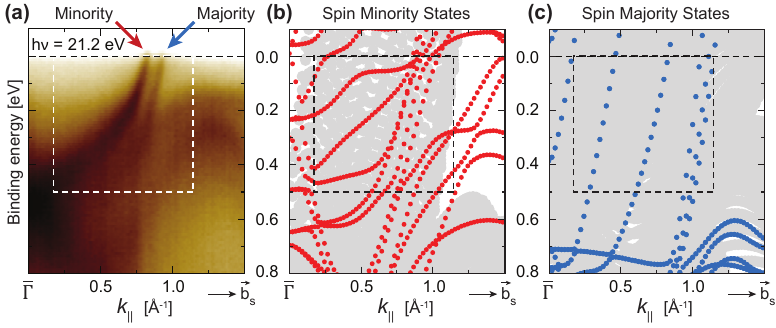}
    \vspace*{-5mm}
    \caption{A comparison of the calculated and measured bandstructure along $\overline{\Gamma}\rightarrow\vb{b}_{\text{s}}$. \textbf{(a)}: Renormalized spin surface resonances measured by ARPES with $h\nu=\SI{21.2}{\eV}$. \textbf{(b)}: Calculated spin minority surface states (red), superimposed on the surface-projected, spin majority bulk states (grey). \textbf{(c)}: Calculated spin majority surface states (blue), superimposed on the surface-projected, spin minority bulk states (grey). Some calculated surface states within the dashed rectangles appear to qualitatively resemble the measured, renormalized surface resonances.}
    \label{fig:figS3}
\end{figure}

In Fig.~\ref{fig:figS3}, the surface spin minority and majority states of Ni(111) -- as calculated from first principles (DFT), are shown along the direction $\overline{\Gamma}\rightarrow\vb{b}_{\text{s}}$ defined in the main text. The surface states of each spin character are shown to overlap with the projected bulk states of the opposite spin character, thus enabling coupling between surface and bulk states via spin-flip scattering \cite{Claessen2004}.

\section*{Modeling the Line Profiles for Electron-Boson Couplings}
\begin{figure}[ht]
    \centering
    \includegraphics[]{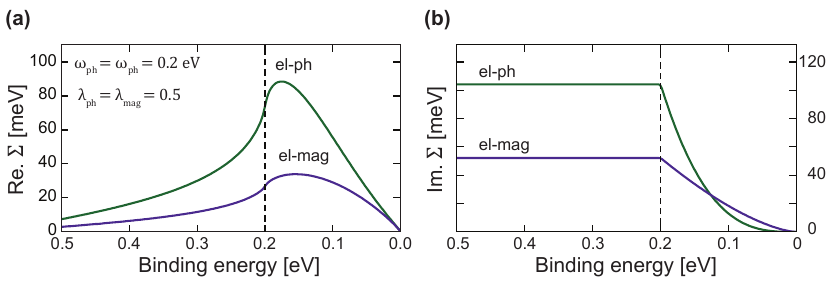}
    \caption{A comparison of the $\Re\Sigma$ and $\Im\Sigma$ of electron-phonon and electron-magnon couplings calculated at $T=\SI{10}{\kelvin}$ using Debye-like models. In both cases $\omega_{\text{max}}=\SI{200}{\milli\eV}$ and $\lambda=0.5$ have been assumed.}
    \label{fig:figS4}
\end{figure}

\vspace{-0.3cm}
In the simple Debye picture, phonons can be described by an energy dispersion $\omega_{\text{ph}}\propto\vqty{\vb{q}}$, a Debye (cut-off) frequency $\omega_{\text{max}}^{\text{ph}}$, and a density of states (DOS) $\rho_{\text{ph}}\pqty{\omega}\propto\omega^{2}$. Under similar assumptions, ferromagnetic magnons will have an energy dispersion relation $\omega_{\text{mag}}\propto\vb{q}^{2}$, a cut-off frequency $\omega_{\text{max}}^{\text{mag}}$, and a DOS $\rho_{\text{mag}}\pqty{\omega}\propto\omega^{1/2}$ \cite{kittel2021introduction}. In the event of electron-boson coupling (EBC) one can -- over a narrow energy and momentum range, assume the electronic DOS to be locally isotropic; and the EBC matrix element $\alpha^{2}$ to be roughly constant with energy and momentum \cite{LaShell2000non,Hellsing2002electron,gayone2005determining}. The effective Eliashberg coupling function $\alpha^{2}F\pqty{\omega}$ is then reduced to the energy-dependent bosonic DOS weighted by the EBC strength. In the simple picture, these can be modeled as \cite{Claessen2004,Hellsing2002electron}:
\begin{align}
    \alpha^{2}F\pqty{\omega}_{\text{ph}}=&
    \begin{cases}
\lambda_{\text{ph}}\pqty{\omega/\omega_{\text{max}}^{\text{ph}}}^{2}, \quad\quad\quad\,& \text{if } \omega\leq\omega_{\text{max}}^{\text{ph}}, \\
        0, & \text{otherwise},
    \end{cases} \label{eq:ph_Debye}\\
    \alpha^{2}F\pqty{\omega}_{\text{mag}}=&
    \begin{cases}
    \pqty{\lambda_{\text{mag}}/4}\pqty{\omega/\omega_{\text{max}}^{\text{mag}}}^{1/2}, & \text{ if } \omega\leq\omega_{\text{max}}^{\text{mag}}, \\
        0, & \text{ otherwise}.
    \end{cases} \label{eq:mag_Debye}
\end{align}
Even with equal cutoff frequencies $\omega_{\text{max}}^{\text{ph}}$, $\omega_{\text{max}}^{\text{mag}}$ and dimensionless coupling strengths $\lambda_{\text{ph}}$, $ \lambda_{\text{mag}}$, the exponent value of $\pqty{\omega/\omega_{\text{max}}}$ in the Eliashberg term will in each case greatly affect the line shapes of the self-energy components $\Re\Sigma$ and $\Im\Sigma$. This is demonstrated in Fig.~\ref{fig:figS4} for both electron-phonon and electron-magnon coupling.

\section*{A Comparison of ARPES-Based Self-Energy Analysis Approaches}

Although simple in form, a Debye model as described will -- in many cases, faithfully recreate the expected line shapes when estimating $\Re\Sigma$ and $\Im\Sigma$. This is especially true for systems with few available bosonic modes, or when certain modes dominate the distinguishable electron-boson interaction \cite{Valla1999many,Mazzola2017strong,mazzola2022tuneable}. When a minimalistic Eliashberg coupling function $\alpha^{2}F\pqty{\omega}$ is assumed, the ARPES-based self-energy analysis will also be less limited by the energy resolution and signal-to-noise ratio of the measurements \cite{chien2009anisotropic}. To determine the electron-boson coupling (EBC) strength $\lambda^{(i)}$ of a specific interaction $i$, one can calculate $\Re\Sigma^{(i)}$ or $\Im\Sigma^{(i)}$ using the model $\alpha^{2}F^{(i)}\pqty{\omega}$ and compare with the experimental self-energy plots. Then $\lambda^{(i)}$ becomes a fitting parameter along with the maximum energy $\omega_{\text{max}}^{(i)}$ of the interaction \cite{hufner2007very,mazzola2022tuneable}. The $\Re\Sigma^{(i)}$ is calculated as \cite{chien2009anisotropic}:
\begin{equation}\label{eq:ReSE}
    \Re\Sigma^{(i)}\pqty{\omega,T} = \hbar\int_{-\infty}^{\infty}\dd\nu\int_{0}^{\omega_{\text{max}}^{(i)}}\alpha^{2}F^{(i)}\pqty{\omega'}\cdot\frac{2\omega'}{\nu^{2}-\omega'^{2}}\cdot f\pqty{\nu+\omega,T}\dd\omega',
\end{equation}
where $f\pqty{\omega,T}$ is the Fermi-Dirac distribution function. $\Im\Sigma^{(i)}$ is calculated using Eq.~4 in the main text. Alternatively, it can be found from $\Re\Sigma^{(i)}$ (and vice versa) via a Kramers-Kronig transformation \cite{Pletikosic2012finding}. If more than one prominent interaction is expected, e.g., a combination of electron-phonon and electron-magnon couplings, a sum of $\Sigma^{(i)}$ contributions with individual $\alpha^{2}F^{(i)}\pqty{\omega}$ can be used \cite{mazzola2022tuneable,rost2023phonon}. Hereinafter, this approach will be referred to as the \emph{Debye method}.

An alternative approach is to make no presumptions about the functional form of the Eliashberg function but instead extract it directly from the experimental $\Re\Sigma$ by inverting the integral in Eq.~\ref{eq:ReSE}. This is typically done using a maximum entropy method (MEM) to overcome the mathematical instability of the inversion problem and prevent overfitting of the experimental data. A detailed outline of the MEM and how it can be applied to ARPES measurements is given in Refs.~\onlinecite{Gubernatis1991quantum,Shi2004direct,tang2004spectroscopic}. Once the Eliashberg function has been extracted from the data the total electron mass-enhancement $\lambda_{\text{tot}}$ can be calculated as \cite{grimvall1981electron}:
\begin{equation}\label{eq:lambda}
    \lambda_{\text{tot}} = 2\int_{0}^{\omega_{\text{max}}}\frac{\alpha^{2}F\pqty{\omega'}}{\omega'}\dd\omega',
\end{equation}
where $\omega_{\text{max}}$ is the maximum observable boson energy.

\begin{figure}[t]
    \centering
    \includegraphics[]{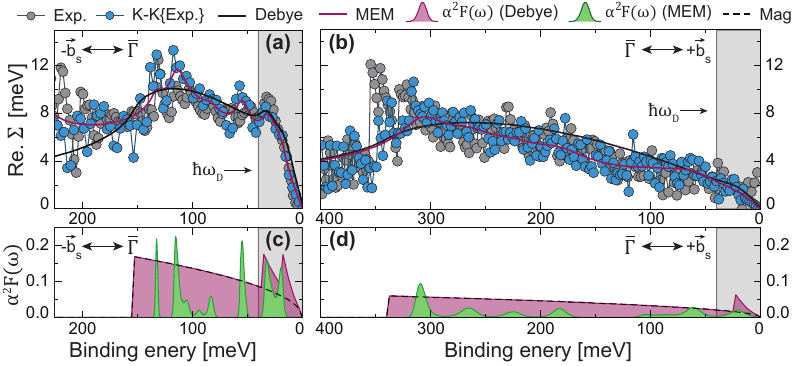}
    \caption{Electron-boson coupling (EBC) in the spin minority surface resonance bands measured within the $1^\text{st}$ PBZ of Ni(111). \textbf{(a,b)}: Kramers-Kronig (K-K) consistent $\Re\Sigma$ of the spin minority bands measured along $\overline{\Gamma}\rightarrow-\vb{b}_{\text{s}}$ (a) and $\overline{\Gamma}\rightarrow+\vb{b}_{\text{s}}$ (b). Both plots have been overlaid by two different models for EBC, one using the Debye method (black lines) and the other using the MEM (purple lines). \textbf{(c,d)}: Model Eliashberg coupling terms $\alpha^{2}F\pqty{\omega}$ for the $\Re\Sigma$ approximations in (a) and (b), respectively. The occupied phonon energy range up to the Debye energy $\hbar\omega_{\text{D}}$ (semi-transparent gray) has been indicated in all four sub-figures.}
    \label{fig:figS5}
\end{figure}

In Fig.~\ref{fig:figS5}, the Debye method and the MEM are both demonstrated on the $\Re\Sigma$ extracted from spin minority bands along directions $\pm\vb{b}_{\text{s}}$ (see the main text, Figs.~2-3). Both data sets have prominent signatures of EBC at energies above and below the Ni Debye energy $\hbar\omega_{\text{D}}$. Either method is seen to reproduce the main features and overall line shape of the measured $\Re\Sigma$. With fewer assumptions and restrictions in place, the MEM yields better overall fits to the data.

The MEM-derived Eliashberg function along $-\vb{b}_{\text{s}}$ (Fig.~\ref{fig:figS5}c, green) exhibits coupling with two distinct Ni phonon modes at binding energies $E_{\text{B}}<\hbar\omega_{\text{D}}$ (semi-transparent gray region). The lowest energy feature ($\approx18$~meV) matches with the Rayleigh mode of the surface \cite{szeftel1985surface,stuhlmann1989surface}. The second feature ($\approx36$~meV) matches with either the bulk longitudinal acoustic or the surface optical mode \cite{Birgeneau1964normal,menezes1990surface}. Along $+\vb{b}_{\text{s}}$ the MEM distinguishes one broad feature ($\approx21$~meV, Fig.~\ref{fig:figS5}d), likely from coupling with one of the bulk transverse acoustic modes \cite{Birgeneau1964normal}. All three electron-phonon couplings (EPCs) can be faithfully recreated by fitting Debye models consisting of one or two separate acoustic modes to the data, as also shown in Fig.~\ref{fig:figS5}.

MEM analysis of the spin minority bands along $\pm\vb{b}_{\text{s}}$ also distinguishes additional electron-boson signatures at larger binding energies up to $\approx320$~meV. Coupling in this energy range is expected to originate from interactions with the bulk magnons of Ni \cite{Cooke1985new,Mook1979neutron,Mook1985neutron,Mook1988temperature,Brookes2020spin}. Based on their energies, the features at $50-150$~meV are thought to originate primarily from coupling to the acoustic magnon dispersion along the $[100]$ direction \cite{Mook1988temperature,Brookes2020spin}; the ones up to $\approx250$~meV primarily from the acoustic $[111]$ branch \cite{Mook1985neutron,Cooke1985new}; and the highest energy couplings from the $[100]$ optical branch \cite{Cooke1985new}. We note that electrons coupling with the optical magnon branch was previously reported from ARPES measurements but never explicitly assigned \cite{Claessen2009}.

Although it will fail to capture the full complexity of $\alpha^{2}F\pqty{\omega}$, a minimalistic acoustic magnon model (Eq.~\ref{eq:mag_Debye}) fitted to the experimental $\Re\Sigma$ of a magnetic system should recreate the expected line shape of the electron-magnon coupling (EMC) region \cite{mazzola2022tuneable}. Furthermore, it should predict the maximum energy of the magnon modes participating in EMC in each case. This is demonstrated in Fig.~\ref{fig:figS5}, where similar overall line shapes and magnon cut-off frequencies have been estimated using the different integrands of the Debye method and the MEM.

Both methods will also -- when optimized, lead to reliable estimates of the total electron-mass enhancement factor \cite{chien2009anisotropic}. The $\lambda_{\text{tot}}$ can be found using Eq.~\ref{eq:lambda}, and the Debye method will also yield $\lambda_{\text{tot}}$ from the sum of the individual fit parameters $\lambda^{(i)}$. Typically, results from the MEM will yield more accurate $\lambda_{\text{tot}}$ values, as fewer simplifications and \emph{ad hoc} assumptions have been made about the shape of the Eliashberg function \cite{Shi2004direct,chien2009anisotropic}. One can also infer contributions to $\lambda_{\text{tot}}$ from the different bosonic modes in cases where one type of EBC is the predominant (or only) coupling mechanism \cite{chien2015electron}. However, this exercise becomes non-trivial when multiple EBC schemes are prominent and their interactions overlap in energy, such as the acoustic phonon and magnon modes of Ni \cite{Birgeneau1964normal,Brookes2020spin}. If a Debye model with multiple modes is used instead, the overlap will be automatically accounted for as the total Eliashberg function will be a sum of individual terms $\alpha^{2}F^{(i)}\pqty{\omega}$ weighted by their coupling strengths $\lambda^{(i)}$. This scenario is exemplified by the best-fit Debye models shown in Fig.~\ref{fig:figS5}. As $\lambda_{\text{tot}}$ is already split into distinct fitting parameters $\lambda^{(i)}$, estimating the individual EBC contributions from the model is straightforward \cite{mazzola2022tuneable}.

\vspace{-0.3cm}
\section*{Additional self-energy plots}
\begin{figure}[t]
    \centering
    \includegraphics[]{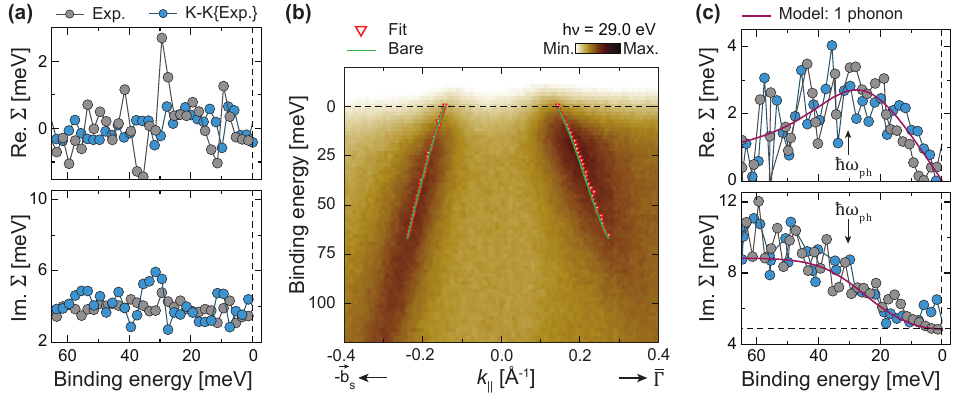}
    \caption{Electron-phonon coupling in the spin minority band along $-\vb{b}_{\text{s}}$.  \textbf{(a)}: The K-K consistent $\Re\Sigma$ and $\Im\Sigma$ of the spin minority resonance at negative $k_{||}$ values in (b). No apparent electron-boson or electron-electron interactions can be observed. \textbf{(b)}: The measured energy bandstructure at $T=\SI{77}{\kelvin}$, overlaid with the one-particle (`Bare', in green) energy band and the experimentally determined, renormalized spin energy band positions (red triangles). \textbf{(c)}: The K-K consistent $\Re\Sigma$ and $\Im\Sigma$ of the spin minority resonance at positive $k_{||}$ values in (b). A single EPC (purple) with $\hbar\omega_{\text{ph}}=\SI{32}{\milli\eV}$ and $\lambda_{\text{ph}}=0.13$, on a background $\Im \Sigma_{\text{el-imp}}+\Delta E_{\text{exp.}}\approx\SI{5}{\milli\eV}$, re-creates the measured $\Sigma$ components.}
    \label{fig:figS6}
\end{figure}

\begin{figure}
    \centering
    \includegraphics[]{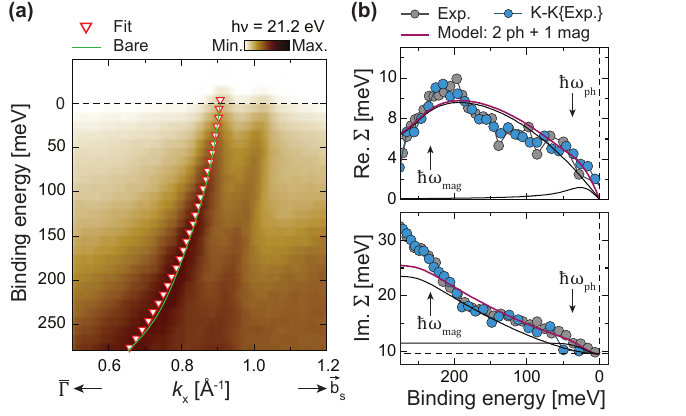}
    \caption{Electron-boson couplings in the spin minority band along $+\vb{b}_{\text{s}}$. \textbf{(a)}: The measured energy bandstructure at $T=\SI{115}{\kelvin}$, overlaid with the one-particle (`Bare', in green) energy band and the experimentally determined, renormalized spin energy band position (red triangles). \textbf{(b)}: The K-K consistent $\Re\Sigma$ and $\Im\Sigma$ of the spin minority band in (a). The renormalizations are best described by two EPCs with a total $\lambda_{\text{ph}}=0.12$ and energies $\hbar\omega_{\text{ph}}^{(1)}=\SI{18}{\milli\eV}$ and $\hbar\omega_{\text{ph}}^{(2)}=\SI{30}{\milli\eV}$, respectively, in addition to one EMC with $\lambda_{\text{mag}}=0.11$ and $\hbar\omega_{\text{mag}}=\SI{250}{\milli\eV}$. A constant energy background $\Im \Sigma_{\text{el-imp}}+\Delta E_{\text{exp.}}\approx\SI{10}{\milli\eV}$ (dashed horizontal line) from electron-impurity scattering and experimental broadening has been assumed.}
    \label{fig:figS7}
\end{figure}

Additional self-energy analysis of the spin minority bands along $\Bar{\Gamma}\rightarrow\mp\vb{b}_{s}$, measured using different photoexcitation energies, is presented in Figs.~\ref{fig:figS6} and \ref{fig:figS7}, respectively. The spin minority surface resonance crossing the Fermi level along $-\vb{b}_{s}$ reveals a distinguishable renormalization only in parts of the measured bandstructure. This was best modeled by one EPC mode with $\hbar\omega_{\text{ph}}=32\pm\SI{2}{\eV}$ and $\lambda_{\text{ph}}=0.13\pm0.02$. The spin minority surface resonance along $+\vb{b}_{s}$ was best modeled by two EPCs, at $\hbar\omega_{\text{ph}}^{(1)}=18\pm\SI{8}{\eV}$ and $\hbar\omega_{\text{ph}}^{(2)}=30\pm\SI{8}{\eV}$ with an added $\lambda_{\text{ph}}=0.12\pm0.04$; and one EMC at $\hbar\omega_{\text{mag}}=250\pm\SI{6}{\eV}$ with $\lambda_{\text{mag}}=0.11\pm0.01$.

While the EPC measured from the spin minority resonances have approximately similar interaction energies $\hbar\omega$ and strengths $\lambda_{\text{ph}}$, the observable EMC can be seen to change dramatically between different positions in the bulk BZ. This is well demonstrated by comparing the $\Re\Sigma$ of the spin minority band near the $\bar{\text{K}}$ and $\bar{\text{M}}'$ points of the projected BZ, but measured using different photon energies $h\nu$ resulting in different $\vb{k}_{\text{z}}$ positions (Figs. \ref{fig:figS5}d vs. \ref{fig:figS7}b). Traversing towards the center of the first zone (i.e., using lower $h\nu$) causes the EMC strength to double, and the maximum observable interaction energy to drop to the middle of the optical branch \cite{Cooke1985new}. The exact reason for this is unclear, but we hypothesize that it occurs because of scattering to/from a different spin majority band, for instance near the $\text{X}$-point of the bulk BZ (see Fig.~1 in Ref.~\onlinecite{Osterwalder2006}).

\begin{figure}
    \centering
    \includegraphics[]{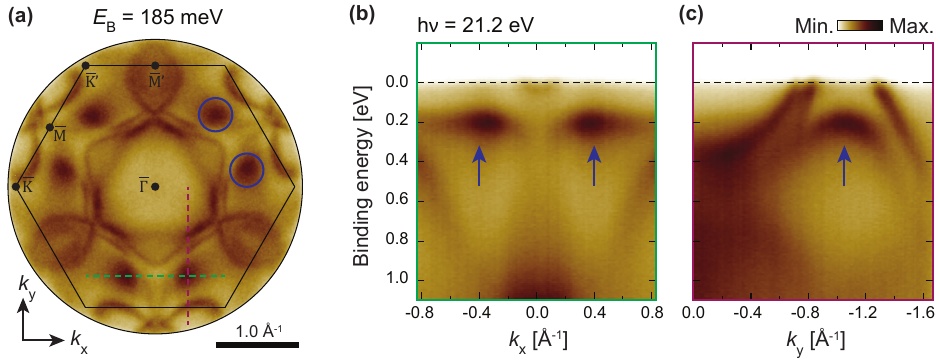}
    \caption{Global energy maximum of the spin majority band from Fig.~4 within the $1^\text{st}$ projected bulk BZ. \textbf{(a)}: A constant energy surface ($k_{\text{x}}$~vs.~$k_{\text{y}}$) of Ni(111) at binding energy $E_{\text{B}}=\SI{185}{\milli\eV}$. \textbf{(b)}: $E_{\text{B}}$~vs.~$k_{x}$, parallel to the $\Bar{\text{K}}'-\Bar{\text{M}}-\Bar{\text{K}}$ line at $k_{y}=-1.1~\text{Å}^{-1}$. \textbf{(c)}: $E_{\text{B}}$~vs.~$k_{y}$, parallel to the $\Bar{\text{M}}-\Bar{\Gamma}-\Bar{\text{M}'}$ line at $k_{x}=0.4~\text{Å}^{-1}$. The maxima are marked with blue circles/arrows.}
    \label{fig:figS8}
\end{figure}

\section*{Global Maximum of the Spin Majority Band in Fig.~4}
Two orthogonal $E$~vs.~$k$ cuts through the maximum of the spin majority band of Fig.~4 in the main text, at $k_{\text{x}}=0$ and $k_{\text{y}}=0$, are shown in Fig.~\ref{fig:figS8}. The two cuts demonstrate that a global maximum for the spin majority band is found at approximately $\pqty{k_{\text{x}},k_{\text{y}}}=(+0.4~\text{Å}^{-1},-1.1~\text{Å}^{-1})$.

%